\documentclass[aps,prl,a4paper,floatfix,showpacs,notitlepage,twocolumn]{revtex4-1} 
\usepackage{amsmath}
\usepackage{amsfonts}
\usepackage{amssymb}
\usepackage{graphicx}
\graphicspath{{./}}
\usepackage{epsfig}
\usepackage{epstopdf}
\usepackage{xcolor}
\usepackage{empheq}
\usepackage{multirow}
\usepackage{cancel}
\usepackage{braket}
\newcommand\nb[1]{\textcolor{blue}{#1}}

\begin{document}
\title{Backaction-evading measurement of entanglement in optomechanics}
\author{Francesco Massel}
\email[]{francesco.p.massel@jyu.fi} 
\affiliation{Department of Physics and
  Nanoscience Center, University of Jyv{\"a}skyl{\"a}, P.O. Box 35 (YFL), FI-40014
  University of Jyv{\"a}skyl{\"a}, Finland}
\date{\today}

\begin{abstract}
  We propose here a fully backaction-evading scheme for the measurement of the
  entanglement between two nanomechanical resonators. The system, which consists
  of two mechanical oscillators, coupled to a single mode of an electromagnetic
  resonant cavity through a radiation-pressure interaction term, is driven by
  two pump tones and four detection tones. As previously discussed in the
  literature, the former induce entanglement between the two mechanical
  oscillators, while we show here that a specific choice of phase and amplitude
  of the detection tones allows for direct pairwise reconstruction of the collective
  quadrature fluctuations of the mechanical oscillators belonging to
  quantum-mechanics-free subspaces, thereby providing
  direct evidence of the entanglement properties of the two mechanical
  resonators.
\end{abstract}
\maketitle

\section{Introduction}
\label{sec:introduction}

The study of the properties of mechanical systems at the nanoscale represents an
extremely active field of research, both at the fundamental and the applied
level
\cite{Maldovan:2013cb,Singh:2014dp,DeAlba:2016bv,ArrangoizArriola:2018gk,Satzinger:2018et}. While
the quantum harmonic oscillator is arguably one of the first examples
encountered in the study of quantum systems, the experimental realization of a
quantum state for a typical ``mass + spring'' mechanical system has achieved only
recently
\cite{OConnell:2010br,Teufel:2011jga,Riedinger:2018aa,OckeloenKorppi:2018ks}. In
the preparation and detection of quantum states for mechanical devices, the
physics of cavity optomechanical systems have played a prominent role
\cite{Aspelmeyer:2014ce}. These systems allow for the manipulation of mechanical
degrees of freedom through a radiation-pressure coupling acting between a
mechanical resonator and an electromagnetic field within a resonant
cavity. Experiments in this sense have led to the preparation of different
quantum mechanical states such as the quantum ground state of a nanodrum coupled
to a microwave cavity \cite{Teufel:2011jga}, quantum-squeezed
\cite{Wollman:2015gx,Pirkkalainen:2015ki,Lecocq:2015dk} and entangled states for
nanomechanical resonators \cite{Riedinger:2018aa,OckeloenKorppi:2018ks}.

One of the crucial aspects concerning the manipulation of quantum mechanical
degrees of freedom --in particular for optomechanical systems-- is represented
by the strategies aimed at certifying, through measurement, that the desired
mechanical state has indeed been prepared.  Within the theoretical framework of
quantum measurement \cite{Wiseman:2010vw} and, more in general, of the physics
of open quantum systems \cite{Gardiner:2004wq}, different strategies have been
proposed to control and probe the state of the optomechanical systems
\cite{Mancini:1998vj,Mancini:2002cn,Vitali:2007ty,Marquardt:2007dn,Teufel:2011jga,Buchmann:2016ic,Rossi:2018bq,OckeloenKorppi:2018hn}.
Of particular relevance to our analysis are the so-called backaction evading
(BAE) measurement setups \cite{Woolley:2013gw,Woolley:2014he}, which aim at
circumventing the effect of the disturbance induced by the measurement apparatus
on the system (backaction) potentially compromising the preparation of a given
quantum state --see, e.g.
Refs. \cite{OckeloenKorppi:2016cp,OckeloenKorppi:2018ks} for recent examples in
this sense.

In this article, we propose a four-tone BAE measurement setup aimed at the
characterization of the entanglement properties of two mechanical resonators.
In particular, the experimental setting we discuss here is constituted by two
mechanical resonators and an electromechanical cavity, either in the optical or
in the microwave regime.

Among the possible measures allowing the quantification of entanglement the most
suitable for our setting is represented by the violation of the Duan bound
\cite{Duan:2000fw}. According to this criterion, to quantify whether such system
is entangled, it is necessary to establish whether the collective quadratures
$X_\Sigma=X_1+X_2$, $Y_\Delta=Y_1-Y_2$ of the mechanical modes violate an
inequality of the form
\begin{align}
  \braket {\Delta X^2_\Sigma}+\braket {\Delta Y^2_\Delta} \geq 1
  \label{eq:1}
\end{align}
where
$\braket{\Delta X^2_\Sigma}=\int \frac{d \omega}{2 \pi} \braket{\left\{X_{\Sigma,\omega},X_{\Sigma,-\omega}\right\}}/2$
and analogously for $\braket {\Delta Y^2_\Delta}$. The operators $X_{1,2}$ and
$Y_{1,2}$ represent the quadrature operators for each of the two mechanical
resonators and fulfil the canonical commutation relations
$\left[X_\mathrm{n},Y_\mathrm{m}\right]=i \delta_{\mathrm{n,m}}$
($\mathrm{n,m}=1,2$). Quadrature operators are proportional to the position
$Q=\sqrt{\hbar/(m \omega_0)} X$ and momentum $P=\sqrt{\hbar m \omega_0} Y$
operators associated with the dynamics of a mechanical oscillator of mass $m$
and resonant frequency $\omega_0$. The goal of our paper is to suggest a
measurement setup allowing for the BAE detection of $\braket {\Delta X^2_\Sigma}$
and $\braket {\Delta Y^2_\Delta} $.

The setting discussed here represents, on the one hand, an improvement over the
detection setup utilized in the experimental verification of the entanglement
between mechanical modes introduced in Ref.~\cite{OckeloenKorppi:2018ks}, in
which $\braket {\Delta X^2_\Sigma}$ could be measured directly --through a BAE
measurement-- whereas $\braket {\Delta Y^2_\Delta} $ was inferred from the
response of the system in the absence of detection probes. On the other hand,
the measurement setup introduced here is an extension of the proposal of
Ref.~\cite{Massel:2017jx}, in which the four-probe setup, while directly
measuring $\braket {\Delta X^2_\Sigma}$ and $\braket {\Delta Y^2_\Delta}$, did
not fulfill the BAE condition, therefore introducing extra backaction noise in
the dynamics of the mechanical resonators, potentially compromising the
mechanical entanglement between the oscillators. 

The paper is organized as follows: after introducing the equations of motion for
the system, we propose a hierarchical solution strategy analogous to the one
introduced in \cite{Massel:2017jx} for the fluctuation operators.  Subsequently,
we show how a specific choice for the probing tones provides a BAE framework
for the detection of mechanical entanglement through the direct measurement of
the output cavity noise spectrum.

In particular, we will show how the current choice of detection tones allows
for the simultaneous BAE measurement of pairs of collective quadratures
belonging to ``quantum-mechanics-free'' (qm-free) subspaces
\cite{Tsang:2012er,Polzik:2015gn,Moller:2017hi} (i.e.~$X_\Sigma$, $Y_\Delta$ or $Y_\Sigma$,
$X_\Delta$). In other terms, we will show that a BAE measurement of either
$X_\Sigma$ or $Y_\Delta$ ($Y_\Sigma$ or $X_\Delta$) will not add any noise to
either quadratures $X_\Sigma$ or $Y_\Delta$ ($Y_\Sigma$ or $X_\Delta$ ),
allowing therefore for a fully BAE detection of the Duan bound.

\section{The system}
\label{sec:system}

\begin{figure}[th!]
  \includegraphics[width=\columnwidth]{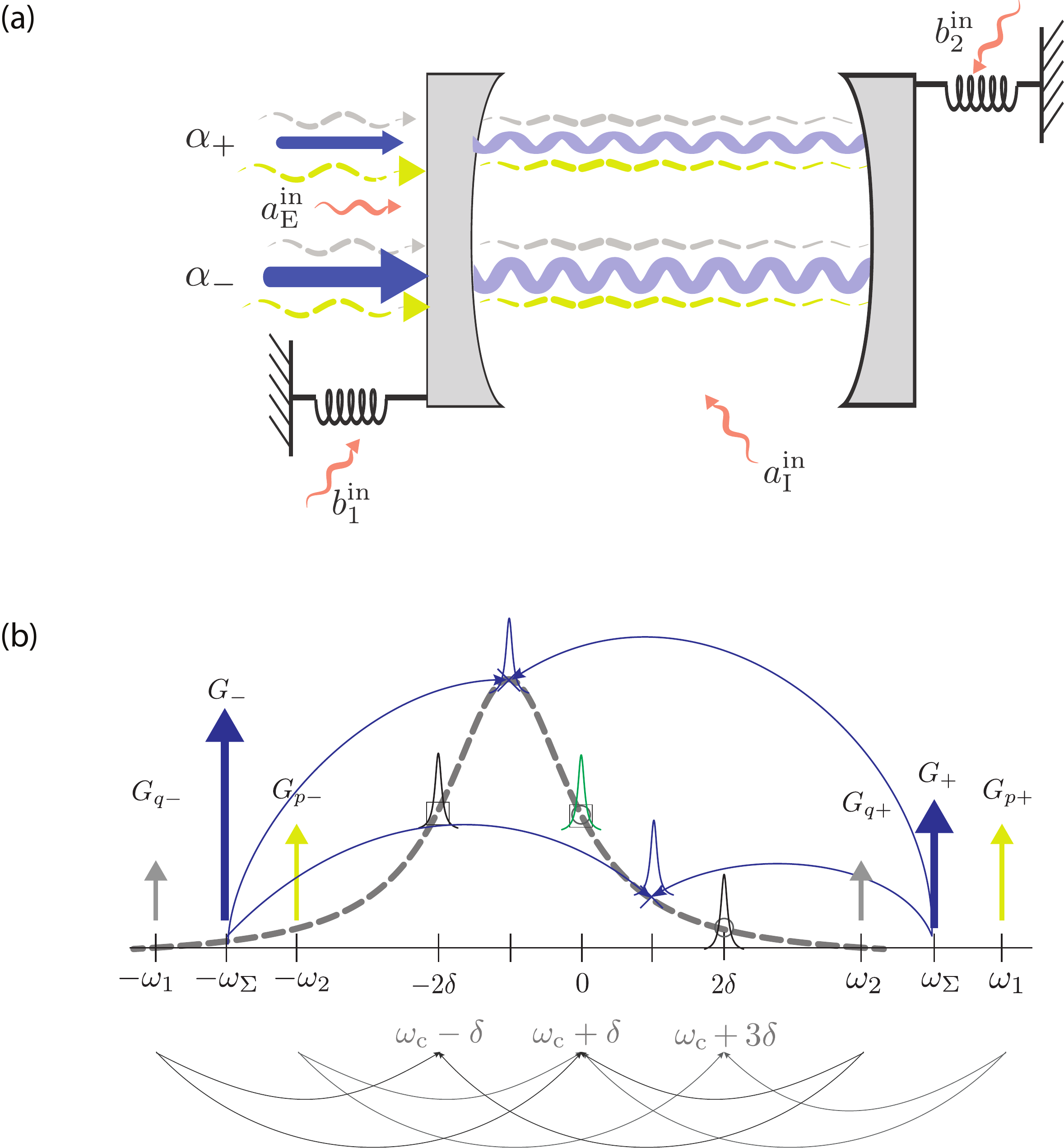}\\
  \caption{(a) Sketch of the proposed entanglement detection scheme. The pump
    tones are depicted as blue (solid dark) arrows, while the detection tones as
    wiggly dashed lines. In addition, we have indicated the three sources of
    noise ($a^\mathrm{in}_\mathrm{I}$,$a^\mathrm{in}_\mathrm{E}$,
    $b^\mathrm{in}$) as short wiggly arrows (b) Pictorial representation of the cavity spectrum
    corresponding to the choice $\omega_+=\omega_1$, $\omega_-=\omega_2$,
    $\omega_\Sigma=\left(\omega_1+\omega_2\right)/2$, $\delta=\left(\omega_1-\omega_2\right)/2$
    (frequency in the rotating frame, see text). The cavity mode is driven with
    two pumps (blue) , generates two sidebands at $\pm\delta$.  In addition to
    the strong driving tone, we consider four probing tones (grey) which
    generate sidebands at $\pm 2\delta$ and $0$. In our analysis we focus on the
    peak generated at 0, which, as we will show contains all the required
    information to ascertain the violation of the Duan bound.}\label{fig:1}
 \end{figure}

The system we are considering consists of a resonant electromagnetic cavity
coupled to two mechanical resonators through a radiation-pressure term. In the
presence of an external coherent field $E(t)$ --denoting with $a$, $b_1$ and
$b_2$ the lowering operators associated with the cavity and the mechanical modes
respectively--, the Hamiltonian for the system can be written as ($\hbar =1$
throughout)
\begin{align}
  H= \omega_{\rm a}  a^\dagger a +
     & \sum_{\mathrm{i=1,2}}\omega_{\mathrm{i}} b_\mathrm{i}^\dagger b_\mathrm{i}
              +g \left(b_\mathrm{i}+b^\dagger_\mathrm{i}\right) a^\dagger a\nonumber\\
     &+i \left[ E(t) a^\dagger - E^*(t) a\right]
  \label{eq:2}
\end{align}
where $\omega_1$, $\omega_2$ and $\omega_\mathrm{a}$ are the resonant
frequencies of the two mechanical oscillators and the cavity, respectively, and
$g$ is the single-photon radiation pressure coupling strength. Furthermore, we
assume that the external field is constituted by a (strong) driving field and a
detection tone $ E(t)=E_\mathrm{drive}(t)+E_\mathrm{detect}(t) $ where
\begin{subequations}
  \begin{align}
    E_\mathrm{drive}(t)=& \alpha_+  e^{-i\omega_{+}t}+\alpha_- e^{-i\omega_{-}t}
                                  \label{eq:3}  \\
    E_\mathrm{detect}(t)=& (\alpha_{\rm p+} e^{i\delta t}+ \alpha_{\rm q+} e^{-i\delta t}) e^{-i\omega_{+}t}\nonumber \\
                                & +(\alpha_{\rm p-} e^{i\delta t}+ \alpha_{\rm q-} e^{-i\delta t}) e^{-i\omega_{-}t}.
                                   \label{eq:4}
  \end{align}
\end{subequations}
As depicted in Fig.~\ref{fig:1}(b), the external field is thus composed of six
tones. Anticipating the results that we will derive below, two of them
($\alpha_+$ and $\alpha_-$ at frequencies $\omega_+$ and $\omega_-$,
respectively) drive the mechanical resonators into an entangled state, while
amplitude and phase of the other four
($\alpha_{\rm p+}$,$\alpha_{\rm q+}$,$\alpha_{\rm p-}$,$\alpha_{\rm q-}$) are
chosen in such a way as to guarantee the BAE measurement of the collective
quadratures ($X_\Sigma$,$Y_\Sigma$,$X_\Delta$,$Y_\Delta$ for symmetric and
antisymmetric modes) of the mechanical resonators. The choice of which
quadrature is being measured, and therefore which quantum-mechanics-free subspace
is being accessed (either $X_\Sigma$, $Y_\Delta$ or $Y_\Sigma$, $X_\Delta$),
depends on the choice of the relative phase between the detection tones 
($\alpha_{\rm p+}$,$\alpha_{\rm q+}$,$\alpha_{\rm p-}$,$\alpha_{\rm q-}$).
\nb{modified here}.

 In order to account for the effect of the environmental noise on the system, we
 introduce a description in terms of quantum Langevin equations (QLEs)
 \cite{Walls:2008em}. Denoting with $\kappa_\mathrm{E}$, $\kappa_\mathrm{I}$ ,
 $\gamma_1$ and $\gamma_2$ the dissipation rates for internal and external
 losses of the cavity and the two mechanical resonators, respectively, we can
 write the QLEs in the interaction frame as
\begin{subequations}
  \begin{align}
    \dot{a}=& -(\kappa/2-i\delta) a -i g \left[(b_1+b_2) e^{-i\omega_{\Sigma}  t} + (b^\dagger_1+b^\dagger_2) e^{i\omega_{\Sigma} t}  \right] a \nonumber \\ 
           & +E(t) e^{i(\omega_a+\delta) t}+\sqrt{\kappa_\mathrm{E}} a_\mathrm{E}^{\rm in}+\sqrt{\kappa_\mathrm{I}} a_\mathrm{I}^{\rm in},\label{eq:5}\\
    \dot{b}_1=& -(\gamma_1/2 + i\delta)b_1 - i g a^\dagger a e^{i \omega_{\Sigma}  t}+ \sqrt{\gamma_1}b^{\rm in}_{1}, \label{eq:6}\\
    \dot{b}_2=& -(\gamma_2/2 - i\delta)b_2 - i g a^\dagger a e^{i \omega_{\Sigma}  t}+ \sqrt{\gamma_2}b^{\rm in}_{2}, 
                \label{eq:7}
  \end{align}
\end{subequations}
where $a_\mathrm{E}^{\rm in}$,$a_\mathrm{I}^{\rm in}$, $b^{\rm in}_{1}$ and
$b^{\rm in}_{2}$ are the input noise operators associated with the coupling of
the cavity to the measurement apparatus ($a_\mathrm{E}^{\rm in}$, external
noise), to the internal losses of the cavity $a_\mathrm{I}^{\rm in}$, and to the
thermal baths coupled to the two mechanical resonators ($b_\mathrm{1}^{\rm in}$
and $b_\mathrm{2}^{\rm in}$), see Fig.~\ref{fig:1}.  The
EOMs~(\ref{eq:5}-\ref{eq:7}) have been written in a rotating frame defined with
respect to
$H_0=\omega_{\Sigma} (b^\dagger_1 b_1 + b^\dagger_2 b_2) + (\omega_{\rm a}+\delta) a^\dagger a$
with $\omega_\Sigma=(\omega_{1}+\omega_{2})/2$,
$ \delta= (\omega_1-\omega_2)/2$, assuming that $\omega_\pm=\omega_{1,2}$.

The noise operators associated with the cavity external and internal losses
$a_\mathrm{i}^{\rm in}$ ($\mathrm{i=I,E}$) obey the relation
$\braket{a_\mathrm{i}^{\rm in}(t)\, a_\mathrm{i}^{\rm in}{^\dagger}(t')}=(n_\mathrm{\mathrm{i}}+1)\delta(t-t')$,
while $b^{\rm in}_\mathrm{j}$ describes thermal noise for the mechanical
resonator and is characterized by the correlation function
$\braket{b_{\rm j}^{\rm in}(t)\,b_{\rm j}^{\rm in \dagger}(t')}=(n_{\rm j}+1)\delta(t-t')$
($\mathrm{j=1,2}$), where $n_{\mathrm{I},\mathrm{E}}$ and $n_{1,2}$ are the
thermal occupation number for the ``internal'' and ``external'' cavity baths,
and each mechanical resonator, respectively.

Following a standard approach, assuming that
$\left|\alpha_\pm\right| \gg \left|\alpha_\mathrm{p \pm}\right|, \left|\alpha_\mathrm{q \pm}\right|$,
Eqs.~(\ref{eq:5}-\ref{eq:7}) can be linearized around the zeroth-order solutions
tones imposed by the driving fields as
\begin{subequations}
  \begin{align}
    &a(t) \to \alpha(t) + a(t) \label{eq:8} \\
    &b_{1,2}(t) \to \beta_{1,2}(t) + b_{1,2}(t) \label{eq:9}.
   \end{align}
\end{subequations}
The corresponding QLEs for the
fluctuations around $\alpha(t)$ and $\beta_{1,2}(t)$ become
\begin{subequations}
  \begin{align}
    \dot{a}=& -(\kappa/2-i\delta) a +\sqrt{\kappa_\mathrm{E}} a_\mathrm{E}^{\rm in}+\sqrt{\kappa_\mathrm{I}}a_\mathrm{I}^{\rm in}  \nonumber \\
            &-i g \left[(b_1+b_2) e^{-i\omega_{\Sigma}  t} + (b^\dagger_1 + b^\dagger_2) e^{i\omega_{\Sigma} t} \right] \alpha \label{eq:10}\\
\dot{b}_1=& -(\gamma_1/2 +i\delta)b_1\nonumber \\
            & - i g \left(\alpha a^\dagger +\alpha^* a\right)e^{i\omega_{\Sigma}t}+ \sqrt{\gamma_1}b_{1}^{\rm in} \label{eq:11}\\
\dot{b}_2=& -(\gamma_2/2 -i\delta)b_2 \nonumber \\
            &- i g \left(\alpha a^\dagger +\alpha^* a\right)e^{i \omega_{\Sigma} t}+ \sqrt{\gamma_2}b_{2}^{\rm in} \label{eq:12}.
  \end{align}
\end{subequations}
where we have assumed that $\delta \gg g\, \beta_{1,2}$. In this case $\alpha(t)$
can be written as
\begin{align}
  \alpha(t)=&\left(\alpha_{+} + \alpha_{\rm p+}  e^{i\delta t} + \alpha_{\rm q+}  e^{-i \delta t}  \right)e^{-i\omega_{\Sigma}t}\nonumber \\
&+ \left(\alpha_{-} + \alpha_{\rm p-}  e^{i\delta t} + \alpha_{\rm q-}  e^{-i \delta t}  \right)e^{i\omega_{\Sigma}t}.
  \label{eq:13}
\end{align}
Assuming that the sideband-resolved condition
($\omega_\mathrm{\Sigma} \gg \kappa$) holds, neglecting terms oscillating at
$\pm 2 \omega_\Sigma$, we can write Eqs.~(\ref{eq:10}-\ref{eq:12}) as
\begin{subequations}
  \begin{align}
    \dot{a}=  &-(\kappa/2 - i\delta) a  +\sqrt{\kappa_\mathrm{E}} a_\mathrm{E}^{\rm in}+\sqrt{\kappa_\mathrm{I}}a_\mathrm{I}^{\rm in}    \nonumber \\
                   &- i \left[\left(G_{-} +G_{\rm p-}e^{i\delta t} + G_{\rm q-}e^{-i\delta t}\right) (b_1 + b_2) \right.\nonumber\\
                   &+\left. \left(G_{+} +G_{\rm p+} e^{i\delta t}+G_{\rm q+} e^{-i\delta t}\right) (b^\dagger_1 + b^\dagger_2) \right] \label{eq:14}\\
    \dot{b}_1=&-(\gamma_1/2 + i \delta)b_1 +\sqrt{\gamma_1}b_{1}^{\rm in} \nonumber \\
                    &  -i \left[ \left(G^*_{-} +G^*_{\rm p-}e^{-i\delta t} + G^*_{\rm q-}e^{i\delta t}\right) a \right.\nonumber \\
                    &+\left. \left(G_{+} +G_{\rm p+}e^{i\delta t} + G_{\rm q+} e^{-i\delta t}\right) a^\dagger \right] , \label{eq:15}\\
    \dot{b}_2=&-(\gamma_2/2 - i \delta)b_2 +\sqrt{\gamma_2}b_{2}^{\rm in} \nonumber \\
                   & -i \left[ \left(G^*_{-} +G^*_{\rm p-}e^{-i\delta t} + G^*_{\rm q-}e^{i\delta t}\right) a \right.\nonumber \\
                  &+\left. \left(G_{+} +G_{\rm p+}e^{i\delta t} + G_{\rm q+} e^{-i\delta t}\right) a^\dagger \right],  \label{eq:16}
   \end{align}
 \end{subequations}
 where $G_{\pm}= g \alpha_{\pm}$ ($G_{\rm q\pm}= g \alpha_{\rm q\pm}$,
 $G_{\rm p\pm}= g \alpha_{\rm p\pm}$) are the linearized optomechanical coupling
 rates associated with the drive and detection tones
 respectively. Eqs.~(\ref{eq:14}-\ref{eq:16}) encode the possibility of
 generating an entangled (two-mode squeezed) state for the two mechanical modes
 by means of the coupling rates $G_\pm$ \cite{Woolley:2014he,Massel:2017jx}. The
 addition of the detection tones $G_{\rm q,p\pm}$ allows for a full
 reconstruction of the collective mechanical quadratures \cite{Massel:2017jx}.
 Most importantly, in contrast to the analysis carried out in
 Ref.~\cite{Massel:2017jx}, and $\alpha_{\rm q \pm}$ can be chosen in such a way
 as to enforce the BAE condition on either the ($X_\Sigma$, $Y_\Delta$) or
 the ($Y_\Sigma$, $X_\Delta$) qm-free subspace. To show this, we introduce a
 Bogolyubov unitary transformation for the mechanical operators
 \begin{align}
    \beta_1= u\, b_1 + v\, b^\dagger_{2} ,\qquad
    \beta_2= u\, b_{2}  + v\, b^\dagger_{1} ,
   \label{eq:17}
  \end{align}
  where $u=G_{-}/ \mathcal{G}$ and $v=G_{+}/ \mathcal{G}$ with
  $\mathcal{G}=\sqrt{G^2_-- G^2_+}$. Without loss of generality, we can assume
  equal mechanical damping rates ($\gamma_1=\gamma_2=\gamma$). In this case, the
  linearized QLE Eqs.~(\ref{eq:14}-\ref{eq:16}) can be written in terms of the
  Bogolyubov modes in the Fourier domain (with the convention
  $a_t \xrightarrow{FT} a_\omega$,
  $a^\dagger_t \xrightarrow{FT} a^\dagger_\omega$) as
\begin{subequations}
  \begin{align}
    \chi^{\rm a ^{-1} }_{\rm \omega+\delta}\, a_{\omega}=
    & -i \mathcal{G} \left[ \beta_{1,\omega} +  \beta_{2,\omega} \right]  +\sqrt{\kappa_\mathrm{E}} a_{\mathrm{E},\omega}^{\rm in}+\sqrt{\kappa_\mathrm{I}}a_{\mathrm{I},\omega}^{\rm in} \nonumber \\
    &-i \left[\mathcal{G}_{\Delta_1}\left\{ \beta_{1, \omega-\delta} +  \beta_{2, \omega-\delta}\right\}  \right.\nonumber \\
    & \quad \quad \left. +\mathcal{G}_{\Delta_3}\left\{\beta^{\dagger}_{1,\omega-\delta} +  \beta^{\dagger}_{2,\omega-\delta}\right\}\right]\nonumber \\
    &-i \left[\mathcal{G}_{\Delta_2}\left\{ \beta_{1, \omega+\delta} +  \beta_{2, \omega+\delta}\right\} \right.\nonumber \\
    & \quad \quad \left.+  \mathcal{G}_{\Delta_4}\left\{\beta^{\dagger}_{1,\omega+\delta} +  \beta^{\dagger}_{2,\omega+\delta}\right\}\right], \label{eq:18} \\
    \chi^{-1} _{\rm \omega-\delta} \,\beta_{\rm 1, \omega}=
    &-i\mathcal{G}    a_{\omega} +\sqrt{\gamma} \beta^{\rm in}_{\rm 1,\omega}\nonumber\\
    & -i  \left[ \mathcal{G}^*_{\Delta_1} a_{\rm \omega+\delta } +\mathcal{G}_{\Delta_3} a^{\dagger}_{\rm \omega-\delta } \right]\nonumber\\
    &-i  \left[\mathcal{G}^*_{\Delta_2} a_{\rm \omega-\delta } +\mathcal{G}_{\Delta_4} a^{\dagger}_{\rm \omega-\delta } \right],
      \label{eq:19} \\
    \chi^{-1}_{\rm \omega+\delta}\, \beta_{\rm 2, \omega}=
    &-i\mathcal{G}    a_{\omega} +\sqrt{\gamma} \beta^{\rm in}_{\rm 2,\omega}\nonumber\\
    &-i  \left[\mathcal{G}^*_{\Delta_1} a_{\rm \omega+\delta } +\mathcal{G}_{\Delta_3} a^{\dagger}_{\rm \omega-\delta } \right]\nonumber\\
    &-i  \left[\mathcal{G}^*_{\Delta_2} a_{\rm \omega-\delta } +\mathcal{G}_{\Delta_4} a^{\dagger}_{\rm \omega+\delta } \right],
      \label{eq:20}
  \end{align}
\end{subequations}
where $\chi^{\rm a}_\omega=\left(\kappa/2-i\omega\right)^{-1}$ and
$\chi_\omega=\left(\gamma/2-i\omega\right)^{-1} $,
$\mathcal{G}_{\rm \Delta_{1,2}}=(u G_{\rm p,q-} -vG_{\rm p,q+})$ and
$\mathcal{G}_{\rm \Delta_{3,4}}=(u G_{\rm p,q+} -vG_{\rm p,q-})$.  Moreover,
$\beta_{\rm 1}^{\rm in}=u b_{\rm 1}^{\rm in} + v b_{\rm 2}^{\rm in ^\dagger}$
and $\beta_{\rm 2}^{\rm in}=u b_{\rm 2}^{\rm in}+v b_{\rm 1}^{\rm in ^\dagger}$
are two correlated thermal noise operators whose only nonzero correlation
functions are
\begin{align}
  \braket{\beta_{\rm 1,\omega}^{\rm in}\, \beta_{\rm 1,\omega'}^{\rm in \dagger}}&=
  \braket{\beta_{\rm 2,\omega}^{\rm in}\, \beta_{\rm 2,\omega'}^{\rm in \dagger}}=
  [(n+1) u^2+n v^2+1]\delta_{\omega+\omega'},
  \label{eq:21} \\
  \braket{\beta_{\rm 1,\omega}^{\rm in \dagger}\, \beta_{\rm 1,\omega'}^{\rm in}}&=
  \braket{\beta_{\rm 2,\omega}^{\rm in \dagger}\, \beta_{\rm 2,\omega'}^{\rm in}}=
  [(n+1) v^2+n u^2]\delta_{\omega+\omega'},
  \label{eq:22} \\
  \braket{\beta_{\rm 1,\omega}^{\rm in }\, \beta_{\rm 2,\omega'}^{\rm in}}&=
  \braket{\beta_{\rm 1,\omega}^{\rm in \dagger}\, \beta_{\rm 2,\omega'}^{\rm in \dagger}}=
  [(2n+1) u v]\delta_{\omega+\omega'},
  \label{eq:23}
\end{align}
where we have assumed the same thermal population for the mechanical resonators
($n=n_1=n_2$).
We now suppose that the probe tones are given by
\begin{align}
  G_{\rm p,q\pm}=G_{\rm p,q}\exp\left[{ \pm i \phi_{\rm p,q}}\right]
  \label{eq:24}
\end{align}
with $G_{\rm p,q}$ real and positive.
In this case, we have that 
\begin{subequations}
  \begin{align}
    \mathcal{G}_{\Delta_1}&=\mathcal{G}_{\rm p} \exp\left[- i\phi_1\right]  \label{eq:25} \\
    \mathcal{G}_{\Delta_3}&=\mathcal{G}_{\rm p} \exp\left[ i\phi_1\right] \label{eq:26}\\
    \mathcal{G}_{\Delta_2}&=\mathcal{G}_{\rm q} \exp\left[- i\phi_2\right]  \label{eq:27} \\
    \mathcal{G}_{\Delta_4}&=\mathcal{G}_{\rm q} \exp\left[ i\phi_2\right] \label{eq:28}
  \end{align}
\end{subequations}
where
$\mathcal{G}_{\rm p, q}=|u e^{-i \phi_{p,q}}-v e^{i\phi_{p,q}}|G_{\rm p,q}$ and
$\phi_{1,2}=\arctan \left[\frac{u+v}{u-v} \tan\left(\phi_{\mathrm{p,q}}\right)\right]$.
With these assumptions, Eqs.~(\ref{eq:18}-\ref{eq:20}) can be solved treating
the probes as perturbations with respect to the pump tones
 \begin{subequations}
   \begin{alignat}{4}
           \mathcal{G}_{\Delta_1}&=\lambda \mathcal{G}_{\rm p} \exp\left[{-i \phi_{\rm 1}}\right], \quad
     && \mathcal{G}_{\Delta_2}&&=\lambda \mathcal{G}_{\rm q} \exp\left[{-i \phi_{\rm 2}}\right],
     \label{eq:29}\\
             \mathcal{G}_{\Delta_3}&=\lambda \mathcal{G}_{\rm p} \exp\left[{i \phi_{\rm 1}}\right],  \quad
     &&   \mathcal{G}_{\Delta_4}&&=\lambda \mathcal{G}_{\rm q} \exp\left[{i \phi_{\rm 2}}\right],
     \label{eq:30}
   \end{alignat}
 \end{subequations}
 where we have introduced the formal perturbative parameter $\lambda$
 ($\lambda=1$ in the end of the calculation). The solution for $a$, $\beta_1$ and $\beta_2$ in
 can be expressed in powers of the perturbative parameter $\lambda$ as
\begin{subequations}
  \begin{align}
    a_{\omega}=&a^{(0)}_{\omega}+\lambda a^{(1)}_{\omega} + \lambda a^{(2)}_{\omega}+ O(\lambda^3), \label{eq:31} \\
    \beta_{1,\omega}=&\beta^{(0)}_{1,\omega}+\lambda \beta^{(1)}_{1,\omega} + \lambda \beta^{(2)}_{1,\omega}+ O(\lambda^3),\label{eq:32}\\
    \beta_{2, \omega}=&\beta^{(0)}_{2,\omega}+\lambda \beta^{(1)}_{2,\omega} + \lambda \beta^{(2)}_{2,\omega}+ O(\lambda^3). \label{eq:33}
  \end{align}
\end{subequations}
Substituting the perturbative expression given in Eqs.~(\ref{eq:31}-\ref{eq:33})
into Eqs.~(\ref{eq:18}-\ref{eq:20})  we get that each term in the perturbative
expansion can be written as
\begin{subequations}
     \begin{align}
       \left(\chi^{\rm a}_{\rm \omega+\delta}\right)^{-1}  a^{\rm (n)}_{\omega}=& -i \mathcal{G} \left[ \beta^{\rm (n)}_{1,\omega} +  \beta^{\rm (n)}_{2,\omega} \right]
                                                                                                         +A_\mathrm{in}^{(\mathrm{n})}, \label{eq:34}  \\
      \left(\chi^\mathrm{m}_{\rm \omega-\delta}\right)^{-1}\beta^{\rm (n)}_{\rm 1, \omega}=&-i\mathcal{G} a^{\rm (n)}_{\omega} +B_\mathrm{1,in}^{(\mathrm{n})}, \label{eq:35} \\
       \left(\chi^\mathrm{m} _{\rm \omega+\delta}\right)^{-1}\beta^{\rm (n)}_{\rm 2, \omega}=&-i\mathcal{G} a^{\rm (n)}_{\omega} +B_\mathrm{2,in}^{(\mathrm{n})}\label{eq:36} 
    \end{align}
  \end{subequations}
  with
  \begin{widetext}
   \begin{subequations}  
  \begin{align}
      A_\mathrm{in}^{(\mathrm{n+1})}=&
        -i \lambda \mathcal{G}_{\rm p}
        \left[
           e^{-i \phi_{1}} \left\{ \beta^{\rm (n)}_{1, \omega-\delta} +  \beta^{\rm (n)}_{2, \omega-\delta}\right\}+
           e^{i \phi_{1}}\left\{\beta^{\rm (n)\dagger}_{1, \omega-\delta} +  \beta^{\rm (n)\dagger}_{2,\omega-\delta}\right\}
       \right] \nonumber \\ &
     -i \lambda \mathcal{G}_{\rm q}
        \left[e^{-i \phi_{2}} \left\{\beta^{\rm (n)}_{1, \omega+\delta} +  \beta^{\rm (n)}_{2, \omega+\delta} \right\}+
             e^{i \phi_{2}}\left\{\beta^{\rm (n)\dagger}_{1, \omega+\delta} +  \beta^{\rm (n)\dagger}_{2,\omega+\delta} \right\}
           \right] \label{eq:40} \\
       B_\mathrm{1,in}^{\mathrm{(n+1)}}=& -i \lambda \mathcal{G}_{\rm p} \left[  
                                                                       a^{\rm (n)}_{\rm  \omega+\delta }    +   a^{\rm (n)\dagger}_{\rm \omega-\delta }\right] e^{i \rm \phi_{1}} 
                                                                       -i \lambda  \mathcal{G}_{\rm q}  \left[  a^{\rm (n)}_{\omega-\delta }           +   a^{\rm (n)\dagger}_{\omega+\delta } \right]e^{i \rm \phi_{2}}, \label{eq:41} \\
       B_\mathrm{2,in}^{\mathrm{(n+1)}}=& -i \lambda \mathcal{G}_{\rm p}\left[  
                                                                       a^{\rm (n-1)}_{\rm  \omega+\delta }    +   a^{\rm (n-1)\dagger}_{\rm \omega-\delta }\right]e^{i \rm \phi_{1}}  
                                        -i \lambda  \mathcal{G}_{\rm q}  \left[  a^{\rm (n)}_{\omega-\delta }           +  a^{\rm (n)\dagger}_{\omega+\delta } \right]e^{i \rm \phi_{2}} \label{eq:42}
  \end{align}
  \end{subequations}
 \end{widetext}
 for $n \geq 0$
 and
 \begin{subequations}
  \begin{align}
     A_\mathrm{in}^{(0)}=&\sqrt{\kappa_\mathrm{E}} a_{\mathrm{E},\omega}^{\rm in}+\sqrt{\kappa_\mathrm{I}}a_{\mathrm{I},\omega}^{\rm in}, \label{eq:37} \\
     B_\mathrm{1,in}^{(0)}=&\sqrt{\gamma} \beta^{\rm in}_{\rm 1,\omega},  \label{eq:38} \\
     B_\mathrm{2,in}^{(0)}=&\sqrt{\gamma} \beta^{\rm in}_{\rm 2,\omega}.
     \label{eq:39}
    \end{align}
  \end{subequations}
Eqs. (\ref{eq:34}-\ref{eq:36}) can be solved to give
 \begin{subequations}
     \begin{align}
       a^{\rm (n)}_{\omega}=&  \frac{\chi^{\rm a}_{\rm \omega+\delta}}{\Delta}
                                 \left\{ A_\mathrm{in}^{(\mathrm{n})}  
                              -i \mathcal{G}
                              \left[
                                   \chi^{\rm m}_{\rm \omega-\delta}    B_\mathrm{1,in}^{\mathrm{(n)}}+
                                   \chi^{\rm m}_{\rm \omega+\delta}    B_\mathrm{2,in}^{\mathrm{(n)}}
                              \right]
                                \right\} \label{eq:43}  \\
       \beta^{\rm (n)}_{\rm 1, \omega}=& \frac{\chi^{\rm m}_{\rm \omega-\delta}}{\Delta}
                                         \left\{ \eta_1 B_\mathrm{1,in}^{\mathrm{(n)}}
                                            -i \mathcal{G}
                                                   \left[ 
                                                     \chi^{\rm a}_{\rm \omega} A_\mathrm{in}^{(\mathrm{n})}
                                                     -i \mathcal{G}  \chi^{\rm m}_{\rm \omega+\delta} B_\mathrm{2,in}^{\mathrm{(n)}}
                                                  \right]
                                         \right\}, \label{eq:44} \\
       \beta^{\rm (n)}_{\rm 2, \omega}=& \frac{\chi^{\rm m}_{\rm \omega+\delta}}{\Delta}
                                         \left\{ \eta_2 B_\mathrm{2,in}^{\mathrm{(n)}}
                                            -i \mathcal{G}
                                                   \left[ 
                                                     \chi^{\rm a}_{\rm \omega} A_\mathrm{in}^{(\mathrm{n})}
                                                     -i \mathcal{G}  \chi^{\rm m}_{\rm \omega-\delta} B_\mathrm{1,in}^{\mathrm{(n)}}
                                                  \right]
                                         \right\}& \label{eq:45} 
    \end{align}
  \end{subequations}
  where
 \begin{subequations} 
  \begin{align}
    \Delta &= 1+ \mathcal{G}^2 \chi^{\rm a}_{\rm \omega+\delta} \left(\chi^{\rm m}_{\rm \omega-\delta} + \chi^{\rm m}_{\rm \omega+\delta} \right) \label{eq:46}\\
    \eta_{\rm 1,2} &= 1+ \mathcal{G}^2 \chi^{\rm a}_{\rm \omega+\delta} \chi^{\rm m}_{\rm \omega \mp \delta} \label{eq:47} 
   \end{align}
 \end{subequations}


For $\mathrm{n}=0$ we have 
($\mathcal{G}_{\rm p}=\mathcal{G}_{\rm q}=0$). 
\begin{subequations}
     \begin{align}
       a^{\rm (0)}_{\omega}=&  \frac{\chi^{\rm a}_{\rm \omega+\delta}}{\Delta}
                              \left\{ \left[\sqrt{\kappa_\mathrm{E}} a_{\mathrm{E},\omega}^{\rm in}+
                                                \sqrt{\kappa_\mathrm{I}}a_{\mathrm{I},\omega}^{\rm in}\right] \right. \nonumber \\
                             & -i \mathcal{G}
                             \left. \left[
                                   \chi^{\rm m}_{\rm \omega-\delta}    \sqrt{\gamma}  \beta_\mathrm{2,\omega}^{\rm in}+
                                   \chi^{\rm m}_{\rm \omega+\delta}    \sqrt{\gamma}  \beta_\mathrm{2,\omega}^{\rm in}
                              \right]
                                \right\} \label{eq:48}  \\
       \beta^{\rm (0)}_{\rm 1, \omega}=& \frac{\chi^{\rm m}_{\rm \omega-\delta}}{\Delta}
                                         \left\{ \eta_1 \sqrt{\gamma}  \beta_\mathrm{1,\omega}^{\rm in}\right.\nonumber \\
                                          &  -i \mathcal{G}
                                                   \left[ 
                                                     \chi^{\rm a}_{\rm \omega}
                                                       \left\{\sqrt{\kappa_\mathrm{E}} a_{\mathrm{E},\omega}^{\rm in}+
                                                                \sqrt{\kappa_\mathrm{I}}a_{\mathrm{I},\omega}^{\rm in}
                                                      \right\}
                                                    \left. -i \mathcal{G}  \chi^{\rm m}_{\rm \omega+\delta} \sqrt{\gamma} \beta_\mathrm{2,\omega}^{\rm in}
                                                  \right]
                                         \right\}, \label{eq:49} \\
       \beta^{\rm (0)}_{\rm 2, \omega}=& \frac{\chi^{\rm m}_{\rm \omega+\delta}}{\Delta}
                                         \left\{ \eta_2 \sqrt{\gamma}  \beta_\mathrm{2,\omega}^{\rm in}\right.\nonumber \\
                                         &   -i \mathcal{G}
                                                   \left[ 
                                         \chi^{\rm a}_{\rm \omega} \left\{\sqrt{\kappa_\mathrm{E}} a_{\mathrm{E},\omega}^{\rm in}+
                                                                                     \sqrt{\kappa_\mathrm{I}}a_{\mathrm{I},\omega}^{\rm in}\right\}
                                                \left.     -i \mathcal{G}  \chi^{\rm m}_{\rm \omega-\delta} \sqrt{\gamma} \beta_\mathrm{1,\omega}^{\rm in}
                                                  \right]
                                         \right\}. \label{eq:50} 
    \end{align}
  \end{subequations}
  
  From these expressions it is possible to see that, if $\gamma \ll \delta$,
  $\beta^{(0)}_{1}$ and $\beta^{(0)}_{2}$ are peaked around
  $\omega \simeq \delta$ and $\omega \simeq -\delta$, respectively, while
  $a^{(0)}$ exhibits a double peak structure for
  $\omega\simeq \pm\delta$. Furthermore, as expected, the solution of
  Eqs.~(\ref{eq:48}-\ref{eq:50}) allows us to establish that the original
  mechanical modes $b_1$ and $b_2$ are entangled, since the cooling of modes
  $\beta_1$ and $\beta_2$ corresponds to two-mode squeezing for $b_1$ and
  $b_2$. Furthermore, we can write the $n=1$ contributions around
  $\omega \simeq 0$ (in the rotating frame) as
\begin{subequations}
  \begin{align}
    a^{(1)}_{\rm \omega}=&-i\frac{\chi^{\rm a}_{\omega+\delta}}{\Delta} \left[\mathcal{G}_{\rm p}
                           \left\{ e^{-i \phi_{1}} \beta^{(0)}_{2,\omega-\delta} +
                           e^{i \phi_{1}} \beta^{(0)\dagger}_{1,\omega-\delta}\right\}\right. \nonumber \\
                         &  +\mathcal{G}_{\rm q} \left. \{e^{-i \phi_{2}} \beta^{(0)}_{2,\omega+\delta} +
                           e^{i \phi_{2}} \beta^{(0)\dagger}_{2,\omega+\delta}\}\right],
                         \label{eq:51} \\
    \beta^{(1)}_{1,\omega}=&0,  \label{eq:52} \\
    \beta^{(1)}_{2, \omega}=&0 \label{eq:53}.
  \end{align}
\end{subequations}
Eq.~(\ref{eq:51}) demonstrates how the dynamics of the mechanical modes
$\beta_1^{(0)}$ and $\beta_2^{(0)}$ can be inferred from the dynamics of the
first-order approximation to the cavity field, and that the measurement is
realized through the presence of the detection tones.

Even though the Bogolyubov operators $\beta_{1,\omega}$ and $\beta_{2,\omega}$
encode all relevant information about the dynamics of the mechanical resonators,
since we are interested in the potential violation of the Duan bound
\eqref{eq:1}, it is more informative to express Eqs.~(\ref{eq:51}-\ref{eq:53})
in terms of frequency-shifted quadrature operators for the collective mechanical
degrees of freedom $\bar{X}^{\Sigma}_{\omega}$, $\bar{X}^{\Delta}_{\omega}$,
$\bar{Y}^{\Sigma}_{\omega}$ and $\bar{Y}^{\Delta}_{\omega}$, defined as
\begin{subequations}
  \begin{align}
    \bar{X}^{\Sigma}_{\omega}=& \bar{X}_{1,\omega}+\bar{X}_{2,\omega},\label{eq:54} \\
    \bar{X}^{\Delta}_{\omega}=& \bar{X}_{1,\omega}-\bar{X}_{2,\omega}, \label{eq:55}
  \end{align}
\end{subequations}
with
$\bar{X}_{1,\omega}=(b_{1,\omega+\delta}+b^\dagger_{1,\omega-\delta})/\sqrt{2}$
and
$\bar{X}_{2,\omega}=(b_{2,\omega-\delta}+b^\dagger_{2,\omega+\delta})/\sqrt{2}$
where analogous definitions hold for $\bar{Y}^{\Sigma}_{\omega}$ and
$\bar{Y}^{\Delta}_{\omega}$.


While for $\delta\neq 0$ original and shifted mechanical quadratures do not
coincide, it is possible to show \cite{Massel:2017jx} that the uncertainties associated with the
shifted mechanical quadratures
$\braket{\Delta \bar{X}^2_{\Sigma}}$  satisfy the following relation
\begin{align}
  \label{eq:56}
\braket{\Delta \bar{X}^2_{\Sigma}}+\braket{ \Delta \bar{Y}^2_{\Delta}}=\braket{\Delta X^2_{\Sigma}}+\braket{ \Delta Y^2_{\Delta}}.
\end{align}
We are thus allowed to express the Duan bound \eqref{eq:1} in terms of
frequency-shifted mechanical quadratures as
\begin{align}
  \braket{\Delta \bar{X}^2_{\Sigma}}+\braket{ \Delta \bar{Y}^2_{\Delta}}\leq 1.
  \label{eq:57}
 \end{align}
 From Eq.~(\ref{eq:51}) and the definiton of the shifted quadratures, it is
 possible to express the first-order correction to the cavity field as
\begin{align}
  a^{\rm (1)}_{\omega}= -i\dfrac{\chi^{\rm a}_{\omega+\delta}}{\sqrt{2}\Delta}
                           &\left[ \left\{ \mathcal{A}^+_{\phi_{\rm p},\phi_{\rm q}}  \bar{X}^{\Sigma}_{\omega}
                                          +\mathcal{B}^+_{\phi_{\rm p},\phi_{\rm q}}  \bar{Y}^{\Sigma}_{\omega} \right\} \right.\nonumber \\
                            & \left.   +i \left\{ \mathcal{B}^-_{\phi_{\rm p},\phi_{\rm q}}  \bar{X}^{\Delta}_{\omega}
                                          - \mathcal{A}^-_{\phi_{\rm p},\phi_{\rm q}}  \bar{Y}^{\Delta}_{\omega} \right\}
                          \right]
     \label{eq:58}                 
\end{align}  
where
\begin{subequations}
  \begin{align}
    \mathcal{A}^{\pm}_{\phi_{\rm p},\phi_{\rm q}}= G_{\rm p} \cos(\phi_{\rm p})\pm G_{\rm q} \cos(\phi_{\rm q})
    \label{eq:59} \\
    \mathcal{B}^{\pm}_{\phi_{\rm p},\phi_{\rm q}}= G_{\rm p} \sin(\phi_{\rm p})\pm G_{\rm q} \sin(\phi_{\rm q}).
    \label{eq:60} 
  \end{align}
\end{subequations}

To ascertain the BAE nature of the current measurement setup in each qm-free
subspace, we need to evaluate the higher-order terms contributing to the cavity
field around the relevant frequency ($\omega \simeq 0$ in the rotating frame).
To this end, from Eqs.~(\ref{eq:40}-\ref{eq:45}), we can write
\begin{align}
   a^{\rm (n+2)}_{\omega}=-i \lambda \frac{\chi^{\rm a}_{\rm \omega+\delta}}{\Delta}
         & \left[
         \mathcal{G}_{\rm p}
         \left\{
                          e^{-i \phi_{1}} \beta^{\rm (n+1)}_{2, \omega-\delta}+
                          e^{i \phi_{1}}\beta^{\rm (n+1)\dagger}_{1, \omega-\delta}
         \right\}\right. \nonumber \\
          &\left.  + \mathcal{G}_{\rm q}
          \left\{
                          e^{-i \phi_{2}} \beta^{\rm (n+1)}_{1, \omega+\delta} +
                          e^{i \phi_{2}} \beta^{\rm (n+1)\dagger}_{2,\omega+\delta}
          \right\}
          \right]
  \label{eq:61}
\end{align}
where, since $\gamma \ll \delta$, we have neglected all non-resonant terms in the
mechanical response $\chi^{\rm a}_{\omega}$.
The terms appearing on the left-hand side of Eq.~\eqref{eq:61} can,
in turn, be expressed as
\begin{subequations}
\begin{align}
  \beta_{\rm 1, \omega +\delta}^{{\rm (n+1)}}=
         -i \lambda \frac{\chi^{\rm m}_{\rm \omega} }{1+ \mathcal{G}^2 \chi^\mathrm{m}_\omega \chi^\mathrm{a}_{\omega+2\delta}}
        & \left[
          \mathcal{G}_\mathrm{p}
               \left( a^{\rm (n)}_{\rm  \omega+2 \delta}    +   a^{\rm (n)\dagger}_{\rm \omega}\right) e^{i \rm \phi_{1}}\right. \nonumber \\
          &\left. + \mathcal{G}_{\rm q}
               \left( a^{\rm (n)}_{\rm \omega} +  a^{\rm (n)\dagger}_{\rm \omega+2 \delta } \right) e^{i \rm \phi_{2}}
            \right] \label{eq:62}\\
    \beta_{\rm 1, \omega -\delta}^{{\rm (n+1)}\dagger}=
         +i \lambda \frac{\chi^{\rm m}_{\rm \omega}}{1+ \mathcal{G}^2 \chi^\mathrm{m}_\omega \chi^\mathrm{a}_{\omega-2\delta}} 
        & \left[
          \mathcal{G}_\mathrm{p}
               \left( a^{\rm (n)}_{\rm  \omega}    +   a^{\rm (n)\dagger}_{\rm \omega-2 \delta}\right) e^{-i \rm \phi_{1}}\right. \nonumber \\
          &\left. + \mathcal{G}_{\rm q}
               \left( a^{\rm (n)}_{\rm \omega-2\delta} +  a^{\rm (n)\dagger}_{\rm \omega } \right) e^{-i \rm \phi_{2}}
            \right] \label{eq:63} \\
    \beta_{\rm 2, \omega -\delta}^{{\rm (n+1)}}=
        -i \lambda \frac{\chi^{\rm m}_{\rm \omega}}{1+ \mathcal{G}^2 \chi^\mathrm{m}_\omega \chi^\mathrm{a}_{\omega}}  
        & \left[
          \mathcal{G}_\mathrm{p}
               \left( a^{\rm (n)}_{\rm  \omega}    +   a^{\rm (n)\dagger}_{\rm \omega+2\delta}\right) e^{i \rm \phi_{1}}\right. \nonumber \\
          &\left. + \mathcal{G}_{\rm q}
               \left( a^{\rm (n)}_{\rm \omega-2\delta} +  a^{\rm (n)\dagger}_{\rm \omega } \right) e^{i \rm \phi_{2}}
            \right] \label{eq:64}\\
    \beta_{\rm 2, \omega +\delta}^{{\rm (n+1)}\dagger}=
          +i \lambda  \frac{\chi^{\rm m}_{\rm \omega}}{1+ \mathcal{G}^2 \chi^\mathrm{m}_\omega \chi^\mathrm{a}_{\omega}} 
        & \left[
          \mathcal{G}_\mathrm{p}
               \left( a^{\rm (n)}_{\rm  \omega+2 \delta}    +   a^{\rm (n)\dagger}_{\rm \omega}\right) e^{-i \rm \phi_{1}}\right. \nonumber \\
          &\left. + \mathcal{G}_{\rm q}
               \left( a^{\rm (n)}_{\rm \omega} +  a^{\rm (n)\dagger}_{\rm \omega+2 \delta } \right) e^{-i \rm \phi_{2}}
            \right] \label{eq:65}
\end{align}
\end{subequations}
 Substituting Eqs.~(\ref{eq:62}-\ref{eq:65}) into Eq.~\eqref{eq:61}, we obtain
\begin{align}
  a^{\rm (n+2)}_{\omega}= \lambda^2 \frac{\chi^{\rm m}_{\rm \omega} \chi^{\rm a}_{\rm \omega+\delta}}{\Delta}
                                 \mathcal{G}_\mathrm{p}\mathcal{G}_\mathrm{q} \left\{e^{i(\phi_1-\phi_2)}-e^{i(\phi_2-\phi_1)}\right\} \nonumber \\
  \left(a^{\mathrm{(n)}}_{\rm \omega -2 \delta }-a^{\mathrm{(n)}}_{\rm \omega + 2 \delta }\right)
  \label{eq:66}
\end{align}
implying that, for $\phi_1-\phi_2=0, \pi$, all terms $a^{\rm (n+2)}_{\omega}$
($\mathrm{n}>0$) are zero for $\delta \ll \kappa$. This condition, combined with
the expression for $a^\mathrm{(1)}_\omega$ given in Eq.~\eqref{eq:58}, allows us
to conclude that a choice of the detection tone phases, that fulfills the
condition $\phi_1-\phi_2=0, \pi$, leads to the faithful mapping onto the cavity
field of the shifted quadrature field selected by the relative phase of the
detection tones.

\begin{figure}[th!]
 \includegraphics[width=\columnwidth]{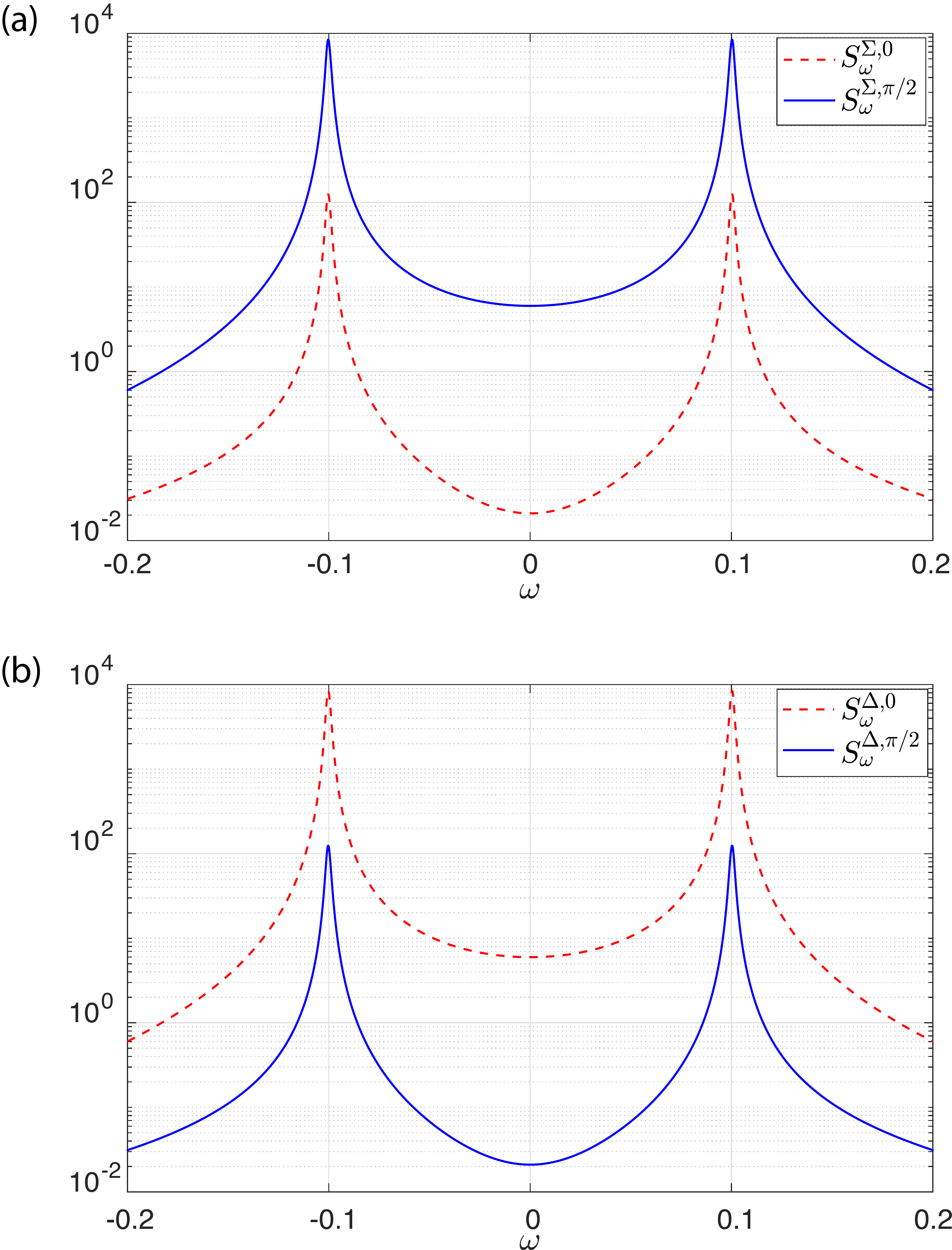}\\
 \caption{ (a) Noise spectrum of the symmetrical mechanical quadrature
   $S^{\Sigma,\theta} _\omega$ as a function of frequency $\omega$ for
   $\theta=0$ (red dotted curve) and $\theta=\pi/2$ (blue solid curve);  (b)
   spectrum of the anti-symmetrical mechanical quadrature
   $S^{\Delta,\theta}_\omega$ for $\theta=0$ (red dotted curve) and
   $\theta=\pi/2$ (blue solid curve). Parameters are $\gamma=10^{-5}$,
   $\delta=0.1$, $G_{-}=4.8\times 10^{-2}$, $G_{+}=4.0\times 10^{-2}$ and
   $n_1=n_2=10$, $n_\mathrm{c}=0$. All frequencies in units of $\kappa$,
   $\hbar=1$ throughout the manuscript.}\label{fig:2}
\end{figure}

 \section{ Spectrum of the output field}

 In the previous section, we have determined that it is possible to access the
 information about the collective dynamics of the mechanical resonators through
 the cavity field $a_\omega$, which does not represent a quantity that is
 directly accessible in experiments. However, through the standard approach
 represented by the I/O formalism \cite{Walls:2008em}, we can relate the cavity
 field to the output field $a^\mathrm{out}_\omega$ --a quantity that can be
 measured in experiments-- as
 $a^{\rm out}_{\omega}=\sqrt{\kappa_{\mathrm E}}a_{\omega}-a_{\mathrm E,\omega}^{\rm in}$.
 To this end, we need to evaluate the expression for the output quadrature field
 in terms of the perturbative expansion given in Eq.~(\ref{eq:31}). Assuming
 that $\phi_1-\phi_2=0,\pi$, we can write the output field quadratures as
\begin{align}
  X^{{\rm out},\theta}(\omega)=&\left[\left(a^{\rm (0) \,out}_{\omega}+
                                              \lambda \sqrt{k_\mathrm{E}} a^{(1)}_{\omega}\right)e^{-i \theta}+ \right.\nonumber \\
                                           &\left.\left({a^{\rm (0) \,out}_{\omega}}^\dagger+
                                             \lambda \sqrt{k_\mathrm{E}}  {a^{(1)}_{\omega}}^\dagger\right)e^{i \theta}\right]
    /\sqrt{2}
  \label{eq:67}
\end{align}
and, since $\braket{a^{\mathrm{(0)}(\dagger)} a^{\mathrm{(1)}(\dagger)}}=0$,
express the spectrum for the output field
$S^{\mathrm{out}}_\omega=\frac{1}{2} \braket{\left\{X^{{\rm out},\theta}(\omega),X^{{\rm out},\theta}(-\omega) \right\}}$
as
\begin{align}
  S^{\mathrm{out}}_\omega=S^{\mathrm{out}(0)}_\omega +\kappa_\mathrm{E} S^{(1)}_\omega
  \label{eq:68}
\end{align}
where
\begin{align}
  S^{\mathrm{out}(0)}_\omega=&\frac{1}{2}\braket{\left\{X^{{(0)\rm out},\theta}(\omega),X^{{(0)\rm out},\theta}(-\omega) \right\}}\nonumber\\
                                     =&\left|\kappa_\mathrm{E} \chi^\mathrm{a}_\omega-1\right|^2\left(n_\mathrm{c}+\frac{1}{2}\right)
  \label{eq:69}
\end{align}
represents the contribution to the output field noise spectrum in the absence of
coupling to the mechanical motion ($\mathrm{G},\, \mathcal{G}_{\rm p,q}=0$).
\begin{figure}[th!]
 \includegraphics[width=\columnwidth]{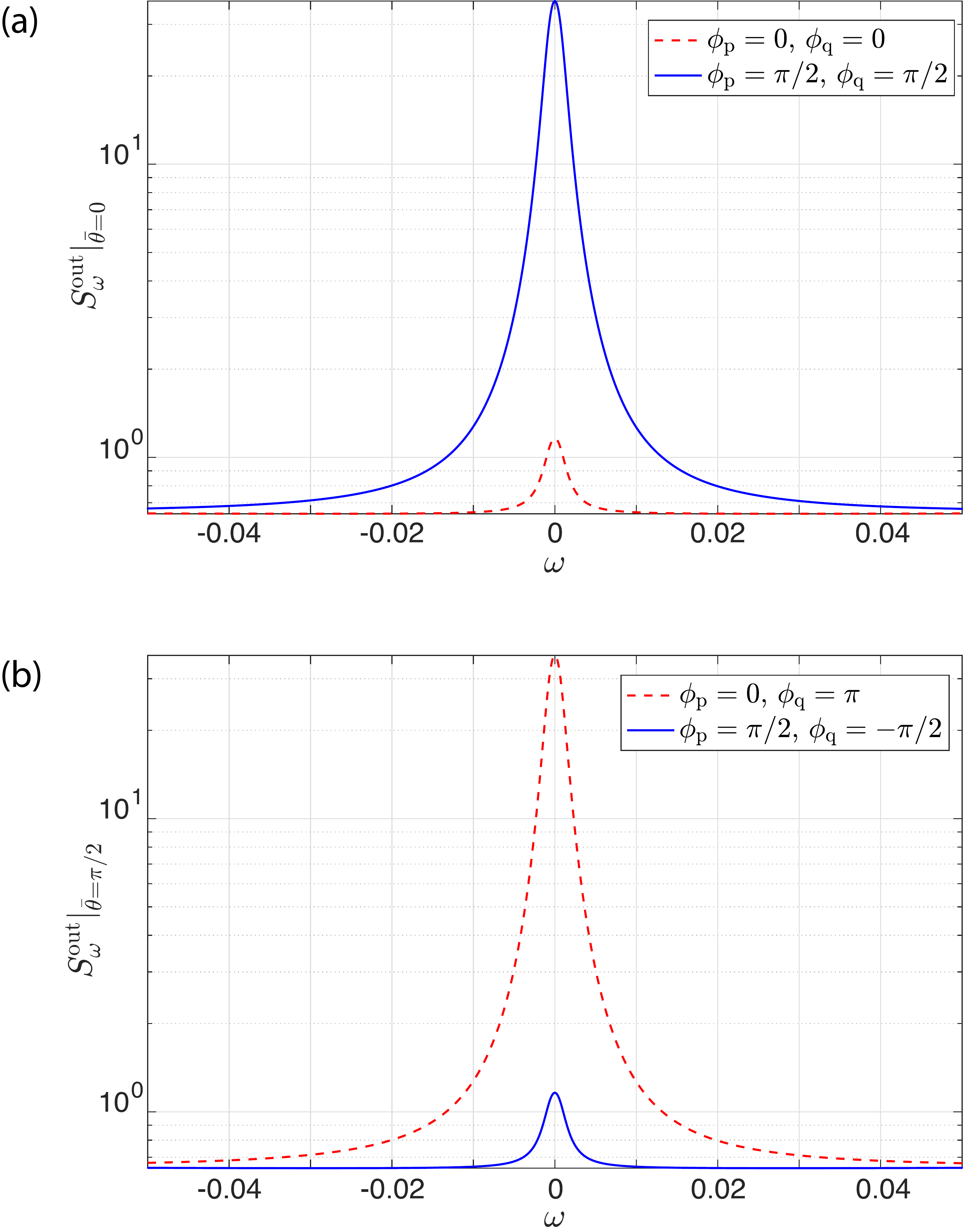}\\
 \caption{(a) Output spectrum 
   $\left. S^{\rm out} _\omega\right|_{\bar{\theta}=0}$ as a function of frequency  for
   $\phi_\mathrm{p}=0,\,\phi_\mathrm{q}=0$ (red dotted curve) and
   $\phi_\mathrm{p}=0,\,\phi_\mathrm{q}=0$ (blue solid curve); (b) Output spectrum 
   $\left. S^{\rm out} _\omega\right|_{\bar{\theta}=\pi/2}$ as a function of frequency  for
   $\phi_\mathrm{p}=0,\,\phi_\mathrm{q}=\pi$ (red dotted curve) and
   $\phi_\mathrm{p}=\pi/2,\,\phi_\mathrm{q}=-\pi/2$ (blue solid
   curve). $\kappa_\mathrm{E}=0.9$, all other parameters as in
   Fig.~\ref{fig:2}. The quantity appearing in the Duan inequality given by
   eq.~\eqref{eq:1} can be inferred from the area under the red dotted curve in
   (a) and the blue solid curve in (b).}
 \label{fig:3}
\end{figure}

More interestingly, $ S^{\rm (1)}(\omega)$ is the contribution to the output
field noise spectrum due to the dynamics of the mechanical oscillators and
therefore represents the relevant term for the determination of a potential
violation of the Duan inequality. From Eq.~(\ref{eq:58}), we have
\begin{align}
  S^{\rm (1)}(\omega)=&\left|\frac{\chi^{\rm a}_{\rm \omega +\delta}}{\Delta}\right|^2 \nonumber \\
                      &\left\{ \cos^2(\theta)\left[{\mathcal{A}^{+}_{\phi_{\rm p},\phi_{\rm q}}}^2 \bar{S}^{\Sigma,0}_{\omega} +
                                                             {\mathcal{B}^{+}_{\phi_{\rm p},\phi_{\rm q}}}^2 \bar{S}^{\Sigma,\pi/2}_{\omega}
                                                      \right]\right.\nonumber \\
                     +&\left.\sin^2(\theta) \left[ {\mathcal{B}^{-}_{\phi_{\rm p},\phi_{\rm q}}}^2 \bar{S}^{\Delta, 0}_{\omega} +
                                                              {\mathcal{A}^{-}_{\phi_{\rm p},\phi_{\rm q}}}^2\bar{S}^{\Delta,\pi/2}_{\omega}
                                                       \right] \right\},
   \label{eq:70}
\end{align}
where
$\bar{S}^{\Sigma,\Delta,0}_{\omega}=\braket{\left\{\bar{X}^{\Sigma,\Delta}_{-\omega},\bar{X}^{\Sigma,\Delta}_{\omega}\right\}}/2$
and
$\bar{S}^{\Sigma,\Delta,\pi/2}_{\omega}=\braket{\left\{\bar{Y}^{\Sigma, \Delta}_{-\omega}, \bar{Y}^{\Sigma, \Delta}_{\omega}\right\}}/2$
are the noise spectra of the frequency-shifted collective mechanical
quadratures, which, upon integration, yield the quantities needed for the
determination of the violation of the Duan bound. From Eq.~(\ref{eq:70}), and
the expressions of $\mathcal{A}^\pm_{\phi_\mathrm{p},\phi_\mathrm{q}}$ and
$\mathcal{B}^\pm_{\phi_\mathrm{p},\phi_\mathrm{q}}$ given in
Eqs.~(\ref{eq:59},\ref{eq:60}), it is clear that the noise spectra of the
collective mechanical quadratures can be accessed from the spectrum of the
output field by changing the phase of the homodyne detector $\theta$ and the
phases of the detection tones $\phi_{\rm p}$ and $\phi_{\rm q}$ (compatibly with
the condition $\phi_1-\phi_2=0,\pi$). In Table \ref{tab:1} we have summarized
the different combinations of ($\bar{\theta}$, $\phi_\mathrm{p}$,
$\phi_\mathrm{q}$) allowing us to access the different frequency-shifted
mechanical spectra, which, upon integration, provide a measurement of the
collective mechanical quadratures needed to ascertain the violation of the Duan
bound.

Furthermore, the choice of $\phi_\mathrm{p}$, $\phi_\mathrm{q}$, in addition to
setting the mechanical quadrature to be measured, fixes the backaction induced
by the measurement tones on it --expressed here as perturbative corrections-- to
be zero. This can be shown by considering the n-th order perturbative term for
the shifted quadrature operators
\begin{subequations}
  \begin{align}
    \bar{X}^{\Sigma,\mathrm{(n)} }_{\omega}&=\frac{u-v}{\sqrt{2}}\left(\beta^{\mathrm{(n)}}_{1,\omega+\delta}+
    \beta^{\mathrm{(n)}\dagger}_{1,\omega-\delta}+
    \beta^{\mathrm{(n)}}_{2,\omega-\delta}+
    \beta^{\mathrm{(n)}\dagger}_{2,\omega+\delta} 
    \right) \label{eq:71} \\
    \bar{X}^{\Delta,\mathrm{(n)}}_{\omega}&=\frac{u+v}{\sqrt{2}}\left(\beta^{\mathrm{(n)}}_{1,\omega+\delta}+
    \beta^{\mathrm{(n)}\dagger}_{1,\omega-\delta}-
    \beta^{\mathrm{(n)}}_{2,\omega-\delta}-
    \beta^{\mathrm{(n)}\dagger}_{2,\omega+\delta} 
    \right) \label{eq:72} \\
    \bar{Y}^{\Sigma,\mathrm{(n)}}_{\omega}&=-i \frac{u+v}{\sqrt{2}}\left(\beta^{\mathrm{(n)}}_{1,\omega+\delta}-
    \beta^{\mathrm{(n)}\dagger}_{1,\omega-\delta}+
    \beta^{\mathrm{(n)}}_{2,\omega-\delta}-
    \beta^{\mathrm{(n)}\dagger}_{2,\omega+\delta} 
    \right) \label{eq:73} \\
    \bar{Y}^{\Delta,\mathrm{(n)}}_{\omega}&=i \frac{u-v}{\sqrt{2}}\left(\beta^{\mathrm{(n)}}_{1,\omega+\delta}+
    \beta^{\mathrm{(n)}\dagger}_{1,\omega-\delta}-
    \beta^{\mathrm{(n)}}_{2,\omega-\delta}-
    \beta^{\mathrm{(n)}\dagger}_{2,\omega+\delta} 
    \right). 
    \label{eq:74} 
  \end{align}
\end{subequations}
which, for $n>0$, represent the backaction contribution to the different
quadrature operators.  Setting $\phi_\mathrm{p}$, $\phi_\mathrm{q}$ in order to
measure a given quadrature (Eq.~\eqref{eq:58}) sets the value of the backaction
contributions to the mechanical quadratures
(Eqs.~(\ref{eq:62}-\ref{eq:65})).

As an example one can choose $\phi_\mathrm{p}=\phi_\mathrm{q}=0$. As it can be
seen from Table~\ref{tab:1}, this choice allows one to measure the
$X^\Sigma_\omega$ quadrature. In turn, substituting the value of the mechanical
Bogolyubov operators from Eqs.~(\ref{eq:62}-\ref{eq:65}) with
$\phi_\mathrm{p}=\phi_\mathrm{q}=0$ into Eq.~\eqref{eq:71}, one can show that
$\bar{X}^{\Sigma,\mathrm{(n+1)} }_{\omega}=0$, demonstrating that the
measurement is backaction evading. At the same time from
Eqs.~(\ref{eq:72}-\ref{eq:74}), the choice $\phi_\mathrm{p}=\phi_\mathrm{q}=0$
also entails that $\bar{Y}^{\Delta,\mathrm{(n+1)} }_{\omega}=0$ --while
$\bar{X}^{\Delta,\mathrm{(n+1)} }_{\omega},\bar{Y}^{\Sigma,\mathrm{(n+1)} }_{\omega}\neq 0$.

Analogous relations hold for the different choices of $\phi_\mathrm{p}$,
$\phi_\mathrm{q}$ giving access, depending on the value of the detection phases,
to the ($X^\Sigma,Y^\Delta$) or the ($X^\Delta,Y^\Sigma$) qm-free
subspace in a fully BAE way.

\begin{figure}[th!]
 \includegraphics[width=\columnwidth]{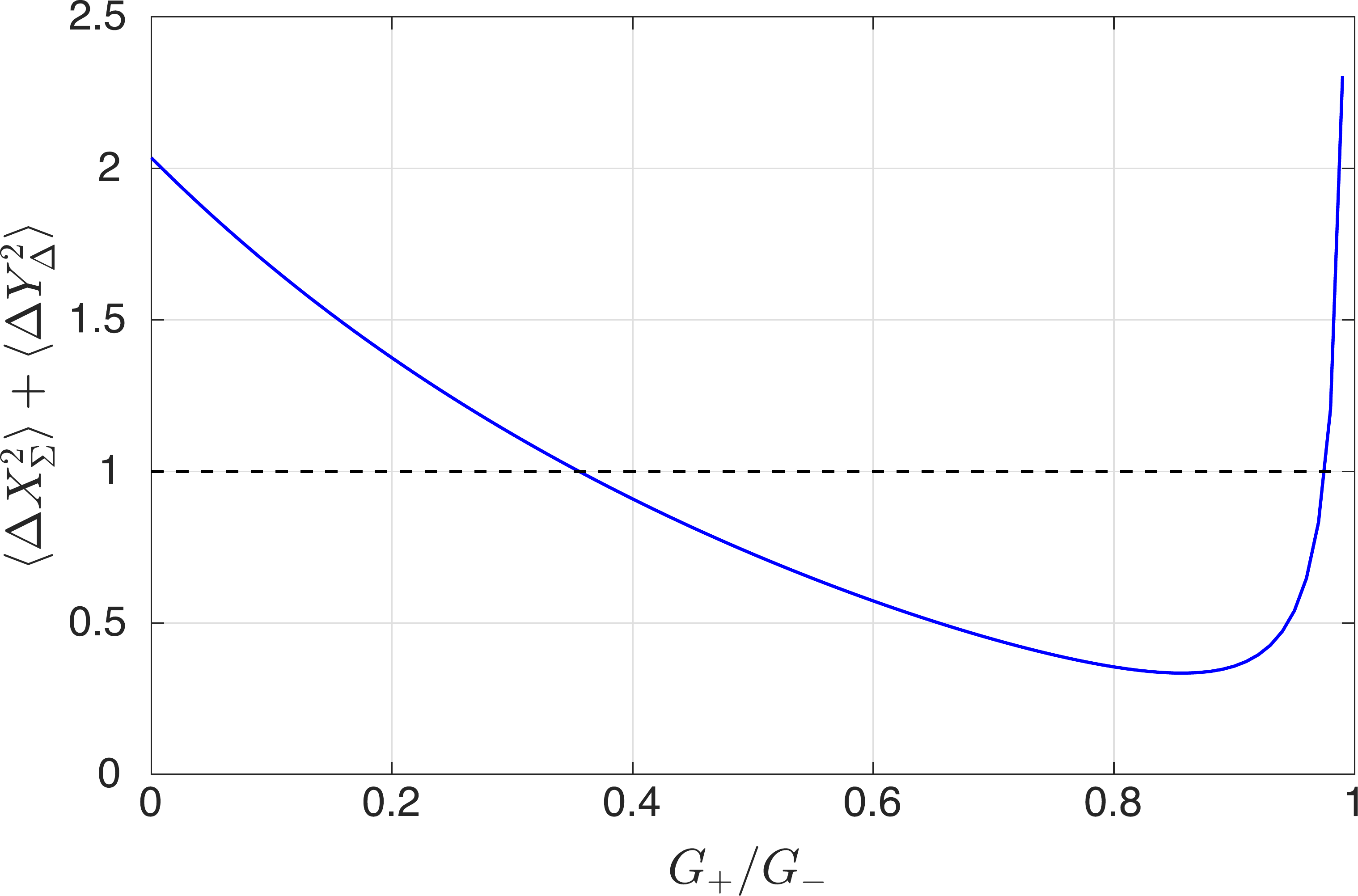}\\
 \caption{ Plot of the Duan quantity in Eq.~(\ref{eq:1}) as a function of ratio
   $G_{+}/G_{-}$. The dashed line indicates the threshold below which the Duan
   inequality is violated. All parameters except $G_+$ as in Figs. \ref{fig:2}
   and \ref{fig:3}. As discussed in the text, the value of the Duan quantity can
   be extracted from the output spectrum as the sum of the integral under the
   red dotted curve in Fig. \ref{fig:3}a and the blue solid curve in
   Fig. \ref{fig:3}b. See also Table~\ref{tab:1}. The values of
   $\phi_\mathrm{p}$ and $\phi_\mathrm{q}$ chosen here imply in all cases that
   $\phi_1-\phi_2=0,\pi$, since
   $\phi_{1,2}=\arctan \left[\frac{u+v}{u-v} \tan\left(\phi_{\mathrm{p,q}}\right)\right]$}
 \label{fig:4}
\end{figure}
\begin{table}[ht]
   \centering
  \begin{tabular}[c]{|l|c|l|}
    \hline
    \multirow{3}{*}{1}
    &$\bar{\theta}=0$
    &\multirow{3}{*}{$S^{\rm (1)}(\omega)=|\frac{\chi^{\rm a}_{\rm \delta}}{\Delta}|^2\left(G_\mathrm{p}+G_\mathrm{q}\right)^2 \bar{S}^{\Sigma,0}_{\omega}$} \\
    &$\phi_\mathrm{p}=0$ & \\
    &$\phi_\mathrm{q}=0$ & \\
    \hline
    \multirow{3}{*}{2}
    & $\bar{\theta}=0$
    &\multirow{3}{*}{$S^{\rm (1)}(\omega)=|\frac{\chi^{\rm a}_{\rm \delta}}{\Delta}|^2\left(G_\mathrm{p}+G_\mathrm{q}\right)^2 \bar{S}^{\Sigma,\pi/2}_{\omega}$} \\
    &$\phi_\mathrm{p}=\pi/2$ & \\
    &$\phi_\mathrm{q}=\pi/2$ & \\
    \hline
    \multirow{3}{*}{3}
    & $\bar{\theta}=\pi/2$
    &\multirow{3}{*}{$S^{\rm (1)}(\omega)=|\frac{\chi^{\rm a}_{\rm \delta}}{\Delta}|^2\left(G_\mathrm{p}+G_\mathrm{q}\right)^2 \bar{S}^{\Delta,\pi/2}_{\omega}$} \\
    &$\phi_\mathrm{p}=0$ & \\
    &$\phi_\mathrm{q}=\pi$ & \\
    \hline
    \multirow{3}{*}{4}
    & $\bar{\theta}=\pi/2$
    &\multirow{3}{*}{$S^{\rm (1)}(\omega)=|\frac{\chi^{\rm a}_{\rm \delta}}{\Delta}|^2\left(G_\mathrm{p}+G_\mathrm{q}\right)^2 \bar{S}^{\Delta,0}_{\omega}$} \\
    &$\phi_\mathrm{p}=\pi/2$ & \\
    &$\phi_\mathrm{q}=-\pi/2$ & \\
    \hline
  \end{tabular}
\caption{Relation between the output spectrum and the shifted mechanical
   quadrature spectra for different values of detection and probe phases.}
  \label{tab:1}
\end{table}

\section{Conclusion}
In this work, we have introduced a 4-probes setup aimed at the measurement of
the entanglement between two mechanical resonators in an optomechanical system,
which is generated by two coherent fields driving the system into a two-mode
squeezed state. We have shown that, if the probing tones are chosen correctly,
within each collective qm-free subspace, no measurement backaction is
present. Furthermore, selecting specific values of the probe phases, the noise
spectrum of each collective mechanical quadrature can be directly mapped onto
the output field noise spectrum. We would like to stress that, while we focused
here on the detection of the entangled state of two mechanical resonators, the
double-BAE detection scheme proposed here is actually independent of the
preparation scheme of the mechanical state, therefore hinting the possibility of
a general BAE characterization of a mechanical systems dynamics within a
quantum-mechanics-free subspace.

\section{Acknowledgements}
The author thanks M. Sillanp\"a\"a and M. Asjad for useful discussions.  This
work was supported by the Academy of Finland (Contract No. 275245).

\end{document}